\documentstyle[12pt]{article}
\def\bra{\left<}
\def\ket{\right>}

\def\I{{\bf I}}
\def\bfq{{\bf q}}
\def\bfh{{\bf g}}
\def\bfQ{{\bf Q}}
\def\bfH{{\bf G}}
\def\bfe{{\bf e}}
\def\bfE{{\bf E}}

\def\d{\partial}

\def\aa{{\mbox{\boldmath$\alpha$}}}
\def\a{{\hbox{\bf H}}}
\def\bfa{{\bf a}}

\def\th{\theta}

\def\F{{\cal{F}}}

\def\de{\delta}

\def\e{\epsilon}

\def\Z{\bf Z}

\def\implies{\Rightarrow}
\def\ie{{\it i.e.,\/}}
\def\eg{{\it e.g.\/}}

\def\qq{\ \ ,\ \ \ }
\def\nl{\nonumber \\ \nonumber \\}

\begin{document}

\newcommand{\inv}[1]{{#1}^{-1}} 

\renewcommand{\theequation}{\thesection.\arabic{equation}}
\newcommand{\beq}{\begin{equation}}
\newcommand{\eeq}[1]{\label{#1}\end{equation}}
\newcommand{\ber}{\begin{eqnarray}}
\newcommand{\eer}[1]{\label{#1}\end{eqnarray}}
\begin{center}
July, 1995 \hfill ITP-SB-95-27 \\
\hfill    RI-6-95 \\
\hfill    hep-th/9508043 \\
\vskip 1.0in
{\large \bf Effective Actions and Gauge Field Stability}\\
\vskip .5in
{\bf Amit Giveon}\footnotemark \\
\footnotetext{e-mail address: giveon@vms.huji.ac.il}
\vskip .1in
{\em Racah Institute of Physics, The Hebrew University\\
  Jerusalem, 91904, ISRAEL} \\
\vskip .15in
and
\vskip .15in
{\bf Martin Ro\v cek}\footnotemark \\
\footnotetext{e-mail address: rocek@insti.physics.sunysb.edu}
\vskip .1in
{\em Institute for Theoretical Physics \\
State University of New York at Stony Brook \\
Stony Brook, NY 11794-3840 USA}\\
\vskip .1in
\end{center}

\vskip .4in
\begin{center} {\bf ABSTRACT } \end{center}
\begin{quotation}\noindent
By studying an effective action description of the coupling of charged gauge
fields in $N=2$ $SU(n)$ supersymmetric Yang-Mills theories, we can describe
regions of moduli space where one or more of these fields becomes
unphysical.  We discuss subtleties in the structure of the moduli
space for $SU(3)$.
\end{quotation}
\vfill
\eject
\setcounter{footnote}{0}
\def\baselinestretch{1.2}
\baselineskip 16 pt
\noindent
\section{Introduction and summary}
\setcounter{equation}{0}

The low energy effective action for the massless fields in $N=2$ supersymmetric
$SU(2)$ Yang-Mills theory was found by Seiberg and Witten in
\cite{sw1,sw2}. These results gave new insight into the dynamics of strongly
coupled theories. In particular, in \cite{sw1} it was shown that there exists a
curve in moduli space across which certain BPS-bound saturated states may cease
to exist.  In \cite{lr}, it was noted that the naive $SU(2)$ covariantization
of the effective action of \cite{sw1} gives a striking signal of this
phenomenon: as one crosses the curve described in \cite{sw1}, the norm of some
states (the charged gauge field supermultiplets) appears to change sign.  If
the theory is to remain unitary, the massive states described by this
covariantized effective action must become unphysical.

In a series of papers \cite{sun,ad,sunq,klt,rest}, the methods of
\cite{sw1,sw2} were extended from the case of $SU(2)$ to general groups.  In
this paper, we consider what we can learn by covariantizing the effective
action and studying where various charged gauge bosons become unphysical. We
give some general results for the Coulomb phase of $N=2$ $SU(n)$ gauge theories
with or without matter hypermultiplets (the generalization to arbitrary simple
groups should be straightforward). We then focus on the case of pure $N=2$
$SU(3)$ gauge theory.  We find that the moduli space is divided into regions
where one or more gauge bosons destabilize.

In more detail, for a general group $G$ with a basis of generators $\{ T_i,
T_\aa\}$, where $\{ T_i\}$ generate the Cartan subalgebra, we write an adjoint
representation $N=1$ chiral superfield $\phi$ as
\beq
\phi^{ab} = A^i(T_i)^{ab}+A^\aa (T_\aa )^{ab}\ \ .
\eeq{A}
Here, $A^i$ describe Abelian superfields, and $A^\aa$ describe charged
superfields.
The low energy effective action for an $N=2$ supersymmetric Yang-Mills theory
in the Coulomb phase
is given in terms of a single holomorphic function $\F (A^i)$ \cite{gates}:
\beq
\frac1{4\pi}Im\left[\int d^4\th\frac{\d\F (\{ A^k\} )}{\d A^i}\bar A^i+
\frac12 \int d^2\th \frac{\d^2\F (\{ A^k\} )}{\d A^i\d A^j}W^{\alpha
i}W_\alpha^j \right] \ \ .
\eeq{sA}
In terms of $\F$, the gauge coupling constants of the theory are given by the
imaginary part of $\tau_{ij}$, where
\beq
\tau_{ij}=\left<\frac{\d^2\F}{\d a_i\d a_j}\right>\ \ ,
\eeq{gc}
and
\beq
a_i=\bra A^i\ket\qq 0=\bra A^\aa\ket\ \ .
\eeq{a}
The expectation values $a_i$ vary as one moves through the moduli space.

We define a gauge-invariant function $\F(\phi^{ab})$ by the condition that it
reduces to $\F (a_i)$:\footnote{This is {\em not} a low momentum expansion of
the 1PI generating functional even at the one-loop level \cite{gr}.}
\beq
\F(\bra\phi^{ab}\ket )=\F (a_i)\ \ ;
\eeq{fphi}
for $G=SU(2)$, $\F(\phi )=\F (\sqrt{\frac12 Tr\,\phi^2})$
\cite{sw1,lr}.\footnote{For general groups, $\F (\phi )$ is a complicated
function of the group invariants; we will not need its explicit form.}
Then a manifestly gauge invariant $N=2$ supersymmetric action \cite{gates}
which reduces to (\ref{sA}) at low energies is
\beq
\frac1{4\pi}Im\left[\int d^4\th\frac{\d\F (\phi )}{\d \phi^{ab}}
(e^V)_{ab,cd}\bar \phi^{cd} +
\frac12 \int d^2\th \frac{\d^2\F (\phi )}{\d \phi^{ab}\d
\phi^{cd}}W^{\alpha ab}W_\alpha^{cd} \right] \ \ ,
\eeq{sphi}
where $V$ is the gauge superfield (see, \eg, \cite{book}). In this work, we
focus on the quadratic terms in (\ref{sphi}); their coefficient is
\beq
\left<\frac{\d^2\F}{\d\phi^{ab}\d\phi^{cd}}\right>\ \ ,
\eeq{coef}
where here, and subsequently,
\beq
\bra f(\phi)\ket \equiv f(\phi)|_{\phi=\bra\phi\ket}\ \ .
\eeq{bra}

In section 2, we compute the explicit form of the generalized gauge couplings
(\ref{coef}) for arbitrary $\F$ in the case of $G=SU(n)$, and find it has a
simple decomposition as a sum of projectors: The terms in the Cartan subalgebra
have $\tau_{ij}$ (\ref{gc}) as coefficients; by construction, these have
positive norm everywhere in moduli space \cite{sw1}.  The remaining terms have
coefficients with imaginary parts that may vanish and change sign in certain
regions of moduli space. As described above, we interpret this phenomenon as a
signal that certain states are destabilizing and disappearing from the
spectrum. We compare this result with the BPS mass formula, and find complete
consistency: As in the $SU(2)$ case \cite{sw1}, when a gauge boson
destabilizes, its mass becomes degenerate with a monopole-dyon pair with the
same total quantum numbers.

In section 3, we analyze the case of $SU(3)$ in more detail. We find a puzzle,
and offer a resolution: certain points in the moduli space should be blown up
into $S_2$'s.\footnote{M.\ Douglas suggests that, while this may be true
mathematically, it isn't relevant physically; see Section 3.}

\noindent
\section{$SU(n)$: The calculation}
\setcounter{equation}{0}
We begin by defining our notation:
We will work in the fundamental $n\times n$ matrix representation of $SU(n)$,
and therefore denote a basis for all $n\times n$ matrices by
\beq
(\bfE^{ij})_{ab}=\de^i_a\de^j_b\ \ ; \ \
\eeq{defE}
we give a special name to the diagonal matrices
\beq
(\bfe_i)_{ab}=(\bfE^{ii})_{ab}\qq i=1,...,n\qq
\eeq{defe}
and choose a basis $\a_i$ of the Cartan subalgebra of $SU(n)$ as well as a dual
basis $\a^*_i$:\footnote{The vectors $\aa_i\equiv diag(\a_i)$ defined by the
diagonal matrices $\a_i$, are a basis for the root lattice. Similarly,
$\aa^*_i\equiv diag(\a^*_i)$ is a basis for the weight lattice. We do not
choose a basis of simple roots corresponding to $\a_i=\bfe_i-\bfe_{i+1}$ as
used in, \eg, \cite{ad,klt}, because this leads to
$\a^*_i=\sum_1^i\bfe_j-\frac{i}n \I$, which we find less convenient than
(\ref{defa}).}
\beq
\a_i=\bfe_i-\bfe_n\qq Tr(\a^*_i\a_j)=\de_{ij}\implies\a^*_i=\bfe_i-\frac1n
\I\qq i=1,...,n-1\ .
\eeq{defa}
Here $\I$ is the $n\times n$ identity matrix.
We parametrize the classical expectation values of the field $\phi$
by eigenvalues $a_i$:
\beq
\left<\phi\right>=\sum_1^{n-1}a_i\a_i=\sum_1^na_i\bfe_i\qq
a_n=-\sum_1^{n-1}a_i\ \ .
\eeq{phi}
We now compute the (generalized) coupling matrix for the extended low energy
effective action defined above in terms of the function $\F$:
\beq
\left<\frac{\d^2\F}{\d\phi^{ab}\d\phi^{cd}}\right>
=\tau_{ij}\left<\frac{\d a_i}{\d\phi^{ab}}\right>
\left<\frac{\d a^j}{\d\phi^{cd}}\right> +
a_{Di}\left<\frac{\d^2a_i}{\d\phi^{ab}\d\phi^{cd}}\right>\qq
\eeq{tabcd}
where
\beq
\tau_{ij}=\left<\frac{\d^2\F}{\d a_i\d a_j}\right>\ \ {\rm and}\ \ \
a_{Di}= \left<\frac{\d\F}{\d a_i}\right>\ \ .
\eeq{tij}
We need to find
\beq
(\bfa_i')_{ab}= \left<\frac{\d a_i}{\d\phi^{ab}}\right>\ \ {\rm and}\ \ \
(\bfa_i'')_{ab,cd}= \left<\frac{\d^2a_i}{\d\phi^{ab}\d\phi^{cd}}\right>\ \ .
\eeq{da}
We do this by differentiating the invariants
\beq
u_k=\frac1k Tr(\phi^k)=\frac1k \sum_1^n(a_i)^k\qq
\eeq{u}
and solving the resulting linear equations. Since the $\phi^{ab}$ are
traceless, differentiation acts as
\beq
\frac{\d\phi^{ab}}{\d\phi^{cd}}=\de^a_c\de^b_d-\frac1n\de^{ab}\de_{cd}\ \ .
\eeq{diff}
Using (\ref{diff}), we differentiate $u_k$ (\ref{u}), and find
\beq
\frac{\d u_k}{\d\phi^{ab}}=(\phi^{k-1})_{ba}-\frac1nTr(\phi^{k-1})\de_{ab}
=\sum_1^n(a_i)^{k-1}\frac{\d a_i}{\d\phi^{ab}}\ \ .
\eeq{du}
Taking the expectation value, and using $\left<\phi^k\right>=\sum(a_i)^k\bfe_i$
(which follows from $\bfe_i\bfe_j=\de_{ij}\bfe_j$), we find
\beq
\sum_1^n(a_i)^{k-1}\bfa_i'=
\sum_1^n(a_i)^{k-1}\bfe_i-\frac1n(\sum_1^n(a_i)^{k-1})\I=
\sum_1^n(a_i)^{k-1}\a^*_i\ \ .
\eeq{eqda}
This is clearly solved by
\beq
\bfa_i'=\a^*_i\ \ ;
\eeq{solda}
for consistency, we must check that $\bfa'_n=
-\sum_1^{n-1}\bfa_i'$, and this is indeed the case.
Thus the $\bfa_i'$ are {\em constant} diagonal matrices that span the Cartan
subalgebra, and we have found that, just as for $SU(2)$ \cite{lr}, the term in
the coupling matrix (\ref{tabcd}) proportional to $\tau_{ij}$ is projected onto
the $U(1)^{n-1}$ subgroup of $SU(n)$, \ie\ onto massless fields.

We now turn to the computation of $\bfa''$. We differentiate (\ref{du})
again, and
find:
\ber
\left<\frac{\d^2u_k}{\d\phi^{ab}\d\phi^{cd}}\right>&=&
\sum_{l=0}^{k-2}\left<(\phi^l)_{bc}(\phi^{k-l-2})_{da}\right>\nl
\ &\ &\ \ \ -\frac{k-1}n\left(\left<(\phi^{k-2})_{ba}\right>\de_{cd}
+\left<(\phi^{k-2})_{dc}\right>\de_{ab}\right)\nl
\ &\ &\ \ \ +\frac{k-1}{n^2}(Tr(\phi^{k-2}))\de_{ab}\de_{cd} \nl
&=&\sum_1^n(a_i)^{k-1}(\bfa_i'')_{ab,cd}
+(k-1)\sum_1^n(a_i)^{k-2}(\bfa_i')_{ab}(\bfa_i')_{cd}\ \ . \nonumber \\
\eer{ddu}
{}From (\ref{phi}) and (\ref{solda}), we find
\ber
&\ &\sum_{l=0}^{k-2}\left(\sum_{i=1}^n(a_i)^l(\bfe_i)_{bc}
\sum_{j=1}^n(a_j)^{k-l-2}(\bfe_j)_{da}\hbox{\hfill}\right)\nl
&\ &-\frac{k-1}n\left(\sum_1^n(a_i)^{k-2}(\bfe_i)_{ba}\de_{cd}
+\sum_1^n(a_i)^{k-2}(\bfe_i)_{dc}\de_{ab}\right)\hbox{\hfill}\nl
&\ &+\frac{k-1}{n^2}\sum_1^n(a_i)^{k-2}\de_{ab}\de_{cd}\hbox{\hfill}\nonumber
\eer{no1}
\beq
=\sum_1^n(a_i)^{k-1}(\bfa_i'')_{ab,cd}
+(k-1)\sum_1^n(a_i)^{k-2}(\a^*_i)_{ab}(\a^*_i)_{cd}\ \ .
\eeq{dda1}
Substituting the explicit form of $\bfe_i$ (\ref{defe}) and $\a^*_i$
(\ref{defa}),  this simplifies to
\beq
\sum_{l=0}^{k-2}\sum_{i,j=1}^n(a_i)^l(a_j)^{k-l-2}\de_b^i\de_c^i\de_d^j\de_a^j-
(k-1)\sum_1^n(a_i)^{k-2}\de_a^i\de_b^i\de_c^i\de_d^i
=\sum_1^n(a_i)^{k-1}(\bfa_i'')_{ab,cd}\ ,
\eeq{dda2}
which can be further simplified to give:
\beq
\sum_{l=0}^{k-2}\sum_{i=1}^n\sum_{j\neq i=1}^n(a_i)^l(a_j)^{k-l-2}
(\bfE^{ij})_{ba}(\bfE^{ji})_{dc}
=\sum_1^n(a_i)^{k-1}(\bfa_i'')_{ab,cd}\ .
\eeq{dda3}
To solve this, we observe that
\beq
\sum_{l=0}^{k-2}(a_i)^l(a_j)^{k-l-2}=\frac{(a_i)^{k-1}-(a_j)^{k-1}}{a_i-a_j}\ ,
\eeq{id}
which allows us to rewrite (\ref{dda3}) as
\beq
\sum_{i=1}^n\sum_{j\neq i=1}^n\frac{(a_i)^{k-1}-(a_j)^{k-1}}{a_i-a_j}
(\bfE^{ij})_{ba}(\bfE^{ji})_{dc}
=\sum_1^n(a_i)^{k-1}(\bfa_i'')_{ab,cd}\ .
\eeq{dda4}
This is clearly solved by
\beq
(\bfa_i'')_{ab,cd}=\sum_{j\neq i}^n\frac{(\bfE^{ij})_{ba}(\bfE^{ji})_{dc}
+(\bfE^{ji})_{ba}(\bfE^{ij})_{dc}}{a_i-a_j}\ \ ;\
\eeq{soldda}
as before, we can easily check the consistency condition
$\bfa_n''=-\sum_1^{n-1} \bfa_i''$.
Note that the $\bfa_i''$ project onto $SU(n)/U(1)^{n-1}$ (for $i\neq j$,
$\bfE^{ij}$ are traceless, and hence are generators of $SU(n)$ outside the
Cartan subalgebra), again as in the
$SU(2)$ case \cite{lr}.  Thus our final expression for the coupling constant
matrix (\ref{tabcd}) is
\ber
\left<\frac{\d^2\F}{\d\phi^{ab}\d\phi^{cd}}\right>
&=&\sum_{i,j=1}^{n-1}\tau_{ij}(\a^*_i)_{ab}(\a^*_j)_{cd}
+\sum_{i=1}^{n-1}\sum_{j\neq i}^{n-1}\frac{a_{Di}-a_{Dj}}
{a_i-a_j}(\bfE^{ij})_{ba}(\bfE^{ji})_{dc}\nl
&\ &\
+\sum_{i=1}^{n-1}\frac{a_{Di}}{a_i-a_n}\left((\bfE^{in})_{ba}(\bfE^{ni})_{dc}
+(\bfE^{ni})_{ba}(\bfE^{in})_{dc}\right)\ .\nonumber\\
\eer{solt}

This is our main result.  It is valid whenever the low energy physics is
described by a function $\F$, \ie\ for the Coulomb phase of an $N=2$ $SU(n)$
gauge theory with or without matter multiplets.  What is the physical
consequence of the computation? By construction, $\tau_{ij}$, the coupling that
we have found in the Cartan subalgebra, has an imaginary part that is positive
everywhere in moduli space. On the other hand, the ratios
$(a_{Di}-a_{Dj})/(a_i-a_j)$ and $a_{Di}/(a_i-a_n)$ can have vanishing and even
negative imaginary parts. The real codimension $1$ surfaces where one of these
ratios becomes real split moduli space into regions where the number of
physical charged gauge field supermultiplets is different.

We now consider the relation of our results in (\ref{solt}) to the BPS mass
formula.  In our notation, the electric charges $\bfq$ are vectors on the root
lattice $\bfq=\sum_1^{n-1}q_i\aa_i\equiv diag(\bfQ )$ where
$\bfQ=\sum_1^{n-1}q_i\a_i$, and by the Dirac quantization condition, the
magnetic charges $\bfh$ are vectors on the weight lattice
$\bfh=\sum_1^{n-1}g_i\aa^*_i\equiv diag(\bfH )$,
$\bfH=\sum_1^{n-1}g_i\a^*_i$. Here $q_i,g_i$ are integers, and the Dirac
quantization condition reads $\sum q_ig_i=Tr(\bfQ\bfH )=integer$. In
particular, the charged gauge bosons $W_{ij}$ have vanishing magnetic charges
$g_k(W_{ij})=0$ and electric charges
\beq
q_k(W_{in})=\de_{ik}\qq q_k(W_{ij})=\de_{ik}-\de_{jk}\ ,\ \ i,j,k=1...n-1\ ,
\eeq{wq}
where we define
\beq
W_{ab}\equiv \sum_1^{n-1} W^i(\a_i)_{ab}+\sum_{i\neq
j=1}^nW_{ij}(\bfE^{ij})_{ab}\ .
\eeq{wab}

The eigenvalues $a_{Di}$ defined in (\ref{tij}) parametrize the expectation
values of the dual field $\phi_D$:
\beq
\left<\phi_D\right>=\sum_1^{n-1}a_{Di}\a_i=\sum_1^na_{Di}\bfe_i\qq
a_{Dn}=-\sum_1^{n-1}a_{Di}\ \ .
\eeq{phid}
The mass $M_{\bfq,\bfh}$ of BPS saturated states is given in terms of the
central charge $Z_{\bfq,\bfh}$ by
\ber
M_{\bfq,\bfh}&=&\sqrt2|Z_{\bfq,\bfh}|\nonumber\\
Z_{\bfq,\bfh}&=&Tr(\bfQ\left<\phi\right>+\bfH\left<\phi_D\right>)
=\sum_1^{n-1}(q_i(a_i-a_n)+g_ia_{Di})\ .
\eer{Z}

Comparing the central charge (\ref{Z}) with our generalized gauge coupling
matrix (\ref{solt}), we see that when the imaginary parts of the coefficients
of the charged gauge fields vanish, their mass becomes degenerate with bound
states of dyons (at threshold, \ie\ with no binding energy). This is an
essential consistency check.\footnote{In the presence of massive
hypermultiplets, (\ref{Z}) receives corrections \cite{sw2}, which do not affect
the charged gauge boson masses.}

\noindent
\section{$SU(3)$: The Physics}
\setcounter{equation}{0}

For pure $N=2$ $SU(3)$ gauge theory, the holomorphic function $\F(a_1,a_2)$ has
been described in
detail \cite{ad,klt}; we can use this to extract more explicit information
about the different regions of moduli space. The terms in the effective action
with the gauge field-strength multiplets $W_\alpha$ are (from
(\ref{sphi}), (\ref{solt})):
\ber
&&\frac1{4\pi}Im\left[\int d^2\th \frac12\tau_{ij}W^{\alpha
i}W_\alpha^j\right.\nl
&&\left. +\frac{a_{D1}-a_{D2}}{a_1-a_2}W^{\alpha12}W_\alpha^{21}+
\frac{a_{D1}}{a_1-a_3}W^{\alpha13}W_\alpha^{31}+
\frac{a_{D2}}{a_2-a_3}W^{\alpha23}W_\alpha^{32}\right]\ ,\nonumber \\
\eer{ss3}
where we have used (\ref{wab}) with $n=3$.  Clearly, depending on the phases of
the three ratios in (\ref{ss3}), one, two or three charged gauge bosons (with
their $CPT$-conjugates) may destabilize.  The authors of \cite{sun,ad,klt}
postulate that the massless gauge coupling matrix $\tau_{ij}$ is defined as the
period matrix of the genus 2 hyperelliptic curve
\beq
y^2=(x^3-ux-v)^2-1\qq
\eeq{curve}
where $u,v$ are coordinates on the quantum moduli space that correspond to
$u_2,u_3$, respectively, in the semiclassical domain, and where, without loss
of generality, we have chosen the dynamically generated scale $\Lambda=1$. Then
$a_i$ and $a_{Di}$ can be calculated as contour integrals:
\beq
I_C=\frac1{2\pi i}\oint_C\frac{x(3x^2-u)dx}{y(x)}\qq
\eeq{aad}
where the contour $C$ runs around various homology cycles on the genus 2
surface corresponding to $a_i,a_{Di}$. This turns out to mean that the integral
(\ref{aad}) is evaluated between various roots of $[y(x)]^2=0$. Different
choices of contours and cuts give different $Sp(4,\Z)$ sections; given an
explicit section on which the charged gauge superfields have a local
description, we could draw a map of walls in moduli space across which gauge
bosons destabilize and disappear.  This requires a careful analysis of the
contours and cuts, which we leave to the future. However, using the results of
\cite{sun,ad,klt}, we can describe some of the phenomena we should find, as
well as a puzzle and a possible resolution.

This quantum moduli space of $SU(3)$ admits a natural $\Z_3\times\Z_2$
action. When the roots of $y^2=0$ fall into three pairs with separations much
larger than the scale (in our conventions, 1), one is in a semiclassical or
weakly coupled region of the moduli space. When precisely two roots degenerate,
then one of three $SU(2)$ subgroups becomes strongly coupled (an ``$SU(2)$
vacuum''); there are six ways that this can happen, and they are rotated into
each other by $\Z_3\times\Z_2$. There are also five special points where the
whole $SU(3)$ is strongly coupled: At three, two pairs of roots simultaneously
degenerate to two distinct points; these are $\Z_2$ invariant and rotate into
each other under $\Z_3$ (``$SU(3)$ vacua''). At the remaining two, three roots
all degenerate to a single point; these are $\Z_3$ invariant and are
interchanged by $\Z_2$ (``$\Z_3$ vacua''). The $SU(2)$ vacua are characterized
by the existence of one massless dyon, the $SU(3)$ vacua are characterized by
the existence of two mutually local massless dyons, and the $\Z_3$ vacua are
characterized by the existence of three mutually {\em non}local massless dyons.

Far from the strongly coupled $SU(3)$ region, each $SU(2)$ vacuum should
reproduce the results of \cite{sw1}. That is, we expect to find a curve passing
through (or close to) the two paired $SU(2)$ vacua on which a charged gauge
field destabilizes, and $\Z_3$ to act by permuting the $SU(2)$ vacua, and
correspondingly, the disappearing gauge fields. As one moves toward stronger
$SU(3)$ coupling, this curve sweeps out a cylinder ($S_1\times{\bf C}$); the
six cylinders corresponding to the six asymptotic $SU(2)$ vacua must meet in
some way in the strongly coupled region.

It is straightforward to see how they meet at the $SU(3)$ vacua: At these
vacua, $v=0$ and $u= 3r^2\th^j$, where
\beq
\th=e^{\frac{2\pi i}3}\qq r=2^{-\frac13}\qq
\eeq{rthdef}
and $j=0,1,2$ labels the three different $SU(3)$ vacua. Without loss of
generality, we may choose $j=0$ (the other choices are found simply by a $\Z_3$
rotation). Then the six roots of $y^2=0$ are $-2r,-r,-r,r,r,2r$.  All possible
integrals (\ref{aad}) are real linear combinations of the integrals
\beq
I_1=\int_{-2r}^{-r}\qq I_2=\int_{-r}^r\qq I_3=\int_r^{2r}\ \ .
\eeq{i123}
However, a glance at (\ref{aad}) shows that $I_1=-I_3, I_2=0$.  Thus all the
quantities $a_i,a_{Di}$ are relatively real at the $SU(3)$ vacuum for {\em any}
$Sp(4,\Z )$ section, and we can conclude that, {\em at the SU(3) vacua, all
three charged bosons simultaneously destabilize}.  This is consistent with
\cite{klt}, where $a_i, a_{Di}$ are explicitly calculated near an $SU(3)$
vacuum for some choice of $Sp(4,\Z )$ section.

Our puzzle arises at the $\Z_3$ vacua.  On general principles, if the $\Z_3$
rotates the various charged gauge fields into each other, as it does in the
semiclassical regions and along the $SU(2)$ vacua, at a $\Z_3$ invariant point,
either zero or three charged gauge fields may destabilize. For a broad class of
$Sp(4,\Z )$ sections, including those of \cite{ad,klt}, it appears that at
least one charged gauge field destabilizes; however, as we show below, because
mutually nonlocal dyons are becoming massless at the $\Z_3$ vacua, all three
charged gauge fields {\em cannot} simultaneously destabilize. At the $\Z_3$
vacua, $u=0,v=\pm1$; without loss of generality, we may take $v=1$. Then the
six roots are $0,0,0,1/r,\th/r,\th^2/r$ (recall (\ref{rthdef})).  By comparing
to the semiclassical limit, it is clear that for any $Sp(4,\Z )$ section on
which the charged gauge bosons are local fields, the $a_i$ are integrals from
$0$ to the root $\th^{i-1}/r$. Then looking at (\ref{aad}), we see that
$a_2=\th a_1$. However, whenever $a_1$ and $a_2$ are {\em not} relatively real,
it follows that all three charged bosons destabilize only if
$a_{Di}=c(a_i-a_3)$ for some real $c$. The mass formula (\ref{Z}) then implies
that only mutually local dyons can become massless simultaneously, which
does not occur at the $\Z_3$ vacua \cite{ad}.

A possible resolution of this puzzle seems to be suggested by the work of
\cite{ad}: They study the vicinity of the $\Z_3$ vacua, and find a modular
parameter $\rho$ {\em that survives at the $\Z_3$ vacuum}.  The gauge couplings
of the massless $U(1)$ fields depend on this parameter. This suggests that at
the $\Z_3$ vacua, the coordinates $u,v$ are not good coordinates, and each
$\Z_3$
vacuum should be blown up into an $S_2$ (with coordinate
$\rho$).\footnote{Indeed, the coordinate transformation $\de u,\de v
\to \rho, \e$ of \cite{ad}, $\de u=3\rho\e^2, \de v=2\e^3$, is precisely such a
blowup.} This would resolve our puzzle: depending on where on the $S_2$ one
sits
(which $\rho$), different charged gauge fields destabilize. However, \cite{ad}
do not make this interpretation; they argue that, although blowing up the
$\Z_3$
points looks more natural mathematically, it is not what is seen physically: at
the singularity, the theory becomes conformally invariant, and the measurable
couplings are the $\Z_3$ symmetric ones, not the $\rho$ dependent ones. With
this interpretation, it is not clear how to interpret the multiplets of charged
gauge bosons near the $\Z_3$ vacua.\footnote{We thank M.\ Douglas for a
discussion on this interpretation.}

We close by noting that a similar analysis could be performed for the Coulomb
phase of the $SU(n)$ theory, with and without matter hypermultiplets.

\bigskip

\bigskip

\noindent
{\bf Acknowledgments}

\bigskip

\noindent
We are happy to thank Mike Douglas for an illuminating discussion about the
physics of the $\Z_3$ point. This work is supported in part by the BSF (the
American-Israel Bi-National Science Foundation). AG thanks the ITP at Stony
Brook and MR thanks the Racah Institute for their respective hospitality. The
work of AG is supported in part by the BRF (the Basic Research Foundation) and
by an Alon Fellowship. The work of MR is supported in part by NSF Grant No.\
PHY
93 09888.

\end{document}